# Understanding Nature's Purpose in Starting all New Lives with Compound Growth

-

# New Science for Individual Systems

For ISSS theme – Future Society

Jessie Henshaw, HDS natural systems design science

sy@synapse9.com

Sunday 7/11/21 talk in the ISSS Special Track on the Future Society

Total 14, 538 words – Abstract 525 – MS 12,260 – # Ref's 66

## Table of Contents

## Abstract

We often associate compound growth with the Anthropocene and our overwhelming economic impacts on the Earth. Today our actual choices for future society appear to lie with studying the class of natural systems that first develop by compound growth, referred to here as *new lives,* natural complex adaptive systems (NCAS). It is preserving the organization of *new lives* that growth creates that is the challenge. Here we offer a way of studying the organizational milestones of *new lives* (MNL), teaching alternative paths to follow and informing a natural systems science for *new lives*, such as those of plants, animals, ecosystems, weather systems, civilizations, economies, communities, businesses, cultures, societies, social groups, personal relationships, and even work habits, home, office, and artistic projects, etc.

Our ability to take part in, create, and guide new lives seems learned from nature, as our methods and nature's show the same organizational milestones. That is the biggest discovery and source of hope that careful study might lead us to splendid choices for how to escape our growing world economic crises. At first, it is easiest to recognize the milestones of growth in familiar subjects and apply that to other subjects. For the Anthropocene, the pattern suggests our economic steering is confusing growing income with wealth, as in a tragedy of the commons, and exposes a plausible creative escape. There might also be much confusion and wonder as nature pushes us to abandon fruitless efforts and follow her path to real success. Though mostly hidden in plain sight now, the natural course ahead might rise through the clouds of confusion and release humanity to act in its own best interests.

The primary milestones for *new lives* include 3 critical events that initiate 3 feedback periods for 3 organizational development stages, occurring in 3 environments that together we can call *egg*, *nest*, *world,* or *natural growth*. A key to studying the growth milestones is learning from life experiences, watching the growth of children, personal relationships, and projects to see how to respond to emerging internal and external relationships. The model is universal, based on the first principle of thermodynamics, the conservation of energy. Energy conservation requires *continuity in energy processes*, forcing organizational development to build smooth shapes like the ubiquitous "S" curves as assembly lines for beginning and ending changes of state.

The most critical milestone is the midpoint turn, from *individuation* during compound growth to maturation during a long climax. That generally comes as new lives exhaust their starting resource and leave their protected *egg* to begin a new life of learning in the *nest*! They must then radically adapt and find internal resources to seek new external resources and prepare for the long future. That shift also marks the inflection point in development rates, from initial positive feedback (relative to the floor) to negative feedback (relative to the ceiling). For future society to do it would take more than a technical change. It would seem to require a *well-informed planetary sense of community,* wide recognition of the milestones, and willingness to take the risk. It seems impossible, of course, but new futures are often not visible from the past. We would only need to trust our ability to innovate and born interest in the success of new lives to make it happen.

Electronic supplementary material

Supplementary topics:   http://synapse9.com/ISSS-21/NewSci-IndividSys-supl.pdf (draft)

Figures Talk Slide set:   http://synapse9.com/ ISSS-21/NewSci-IndividSys-talk.pdf (draft)

Credit Author Statement: Funding, Concepts, and Figures by the author. Proofreading by ProofreadingPal and Grammarly.

# 1   Introduction

> *"We know when something tries to grow forever within a healthy living system;*
> *it is a threat to the health of the whole. So, why would we imagine that our*
> *economy would be the one system that can buck this trend?" — Kate Raworth*
> *(2004) TED Talk*

We offer a scientific study of how natural growth develops self-organized little worlds with lives of their own, *individual new lives*. It refers to a broad category of natural complex adaptive systems (NCAS) related to Miller's (1973a, 1973b) and Varela, Maturana, and Uribe's (1974) models of living systems. This study uses observable stages of organizational growth and change as windows on the otherwise hidden internal worlds of growth systems and their environmental relations: a general natural systems science. Examples of *new lives* include individual plants, animals, ecosystems, economies, communities, businesses, cultures, societies, social groups, individual relationships, and varied other organizational tasks and projects of all kinds. There are also more primitive forms of new lives of some interest to consider, such as self-organizing individual thermal, electrical, and chemical systems, like tornados, lightning, and fire.

We introduce methods of testable observation to support a diagnostic method of studying the growth of new lives and their contexts. We focus on each growth process as a whole, as 3 transformational events, 3 following development feedback periods, and organizational processes, in 3 corresponding environments: *egg*, *nest*, & *world*. Figure 5 provides a graphic aid illustrating each stage building upon the last, pictured as milestones along an "S" curve assembly line of increasingly mature stages: a total of twelve milestones forms a complex hierarchy[1]:

*egg {germ, takeoff, individuation} nest {turn-forward, fitting in, maturation}*
*world {release, life, engagement}.*

The term needing more explanation is *turn-forward.* It is a more general neutral term for the event of *birth,* the often particularly challenging transition from compound growth in the self-contained *egg* environment to the first exposure to the wide world and the work of new learning and *maturation* in the more open *nest* environment. New lives are still extremely immature at that point, *fully formed but quite undeveloped,* and having exhausted their compound growth resource and need to find others or perish. Other terms for the *turn-forward* may be used, depending on the context, like the go-ahead, emergence, coming of age, or birth. *Turn-forward* is just a descriptive general term. In architectural work, that stage is called *conceptual design,* when the intentions are fully formed, but the work is still quite undeveloped. It also applies to mental concepts that may express a clear intent but remain undeveloped. People new to an organization or a profession may be called *novices* or *newbies*, referring to the same fully formed but quite undeveloped stage. Emerging plant *sprouts* or *seedlings* are fully formed but quite undeveloped too, organizationally advanced but highly immature, having just emerged from compound growth as they get their first sheltered taste of the real world.

The reason for so carefully choosing terms is that the biggest discovery, and heart of the discussion, is how similar natural and human methods of creating new lives are. Both display the same developmental milestones. Thus, for example, home and office projects, large and small, start with some seed idea, a *germ* that first develops as a concept for doing things, going through a process of *individuation* in someone's mind. When the concept is fully formed and ready to develop, the focus *turns forward* to making specific plans and preparations for the future by *maturing* the idea in its *nest* environment. Once fully developed, the plan is *released* into the world. That release may result in an enduring steady-state or be the *germ* of a second *egg, nest, world* process on a larger scale.

At first, these new terms may feel a bit unfamiliar. Our normal work of creating and guiding *new lives* is mostly non-verbal, and our responses to the milestones intuitive. The new terms are for enabling discussion of what may often already be familiar. We raise children, organize and enjoy dining and travel events, develop careers, arrange home and office projects, guide businesses, participate in movements, learn new subjects, and build personal and professional relationships. We also try to steer our governments and the world. Each is an organizational effort likely to be better understood by studying its growth milestones. To start, first, identify some process of change or experience of interest. Then see if you can identify the *enclosed egg* it started in, the *sheltered nest* it matured in, and the open *world* in which it was or will be *released*. After trying a few familiar subjects, the terms should fall into place.

---

[1] Note that the terms for the twelve growth milestones are selected for generality. More specific terms may fit better in specific cases. There are also *new lives* that do not seem to fit the pattern, like new cells born by cell division.

The detailed growth model for *new lives* is in §3, and a further introduction to the general theory is in sections §4 and §5. The above brief sketch is enough introduction, though, to follow a "core pedagogy," starting in section §2 with a deep dive into how to relieve the world economy of its threatening fixation on endless compound growth, mostly compounding our problems in its present state. That fixation may be the principal barrier to our ability to respond to natural limits and thrive in the natural world. We use a simple model of the system's basic financial plumbing to illustrate the coupling of finance and commerce. Those represent two main kinds of financial circulation that are increasingly out of balance: the circular flow of money in commerce and the exponential concentration of money in finance. There are also signs of a natural turn-forward already beginning culturally but blocked. Just seeing the problem does not offer an immediate fix, of course, as the current system is deeply engrained. It does permit wider study and comparison to how less troubled systems handle it, though, where the keys to possible splendid solutions would lie.

The change needed in the world economy is like the natural change in many family businesses as they mature. Having long put most of the profits into growing the business, once they succeed, they can switch to using the profits for personal, family, and community needs. That way of climaxing growth for a thriving system can be very smooth. The world economy is also a family business, for a big family having achieved a big but very troubled success now a threat to itself. Part of the problem is our not seeing humanity and nature as family to care for as we become able. So, we have big family decisions to make and need a common language, part of which the *milestones of new lives* might offer.

## 2 Steering for Home

### 2.1 The Economic Challenge

How misguided the growth of the world economy has become is clear from the growing number, variety, and scale of its impacts on the Earth and humanity, threatening our future (Bradshaw et al. 2020; Rees 2020; Steffen et al. 2015, 2016; Meadows et al. 1972, 2004). Institutional research reports on systemic

risks[2],[3] display the same broad pattern, termed the Anthropocene's *great acceleration*,[4] of humanity's ever-escalating disruption of the Earth. An experimental list of the Top 100 World Crises[5] adds less studied systemic societal and economic threats caused by how growth forces people to reorganize how they live faster and faster. Organizational dysfunction and failure are the ultimate limits to growth, sadly left out of other threat assessments. We might call that threat of excess acceleration of change *systemic jerk*, referring to the mathematical name of exponential growth's 3rd derivative, also exponential. That can make growth limit symptoms debilitating, like worsening systemic interference and congestion, growing cultural miscommunication, and failing efforts to cooperate. These threats are harder to measure than material impacts like CO2, of course, but faster. If unrestrained, they would catch us even more off guard and be as or more disruptive.

Several past civilizations have rapidly risen, as we have, but then collapsed, such as Rome (Tainter 1988; Diamond 2005; Lent 2017). Garett Hardin seems to have been the first person to identify the real reason civilizations sometimes seem to aggressively destroy themselves, called the Tragedy of the Commons (1968). People managing resources fall into the trap of mistaking growing income for wealth. As a result, they keep growing their incomes as it depletes their real wealth. However one reads history, the model perfectly fits our situation, with our global financial imperative to maximize income growth. That systemic force easily overwhelms sustainability in an increasingly competitive world. In Hardin's model, village milk cows graze on a shared meadow, and the farmers find they need to put more and more cows out to pasture to get the same amount of milk. The grass and the cows, of course, both wither and die.

No doubt Hardin's villagers felt trapped not knowing why, as we do, not understanding why life seems to be an ever-increasing struggle as on paper we are becoming ever richer. Both are true, and it is related to how we respond to amplifier feedback. At first, it does not bother us; then, we try to ignore it until it becomes painful. The one available remedy is to find the relief valve and turn down the feedback, understanding that nature is full of examples *new lives* of many kinds that solve it quite simply.

---

[2] **2019 WEF Global Risks Report** http://www3.weforum.org/docs/WEF_Global_Risks_Report_2019.pdf
"Global Risks out of Control - Is the world sleepwalking into a crisis? Global risks are intensifying but the collective will to tackle them appears to be lacking. Instead, divisions are hardening. The world's move into a new phase of state-centred politics, noted in last year's Global Risks Report, continued throughout 2018. The idea of 'taking back control'—whether domestically from political rivals or externally from multilateral or supranational organizations—resonates across many countries and many issues. The energy now being expended on consolidating or recovering national control risks weakening collective responses to emerging global challenges. We are drifting deeper into global problems from which we will struggle to extricate ourselves".

[3] **2019 UN Global Assessment Report on Disaster Risk** - https://gar.unisdr.org/ Conclusion: - "Disaster risks emanate from development pathways, manifesting from the trade-offs inherent in development processes. In some ways, this has always been well recognized. What is new in today's increasingly interconnected society is the diversity and complexity of threats and hazards, and the complex interaction among them, which result in "an unprecedented global creation of risks, often due to previous socioeconomic development trends interacting with existing and new development dynamics and emerging global threats." P 418

[4] **The Anthropocene** – documented by the Intnl. Geosphere-Biosphere Program
http://www.igbp.net/globalchange/greatacceleration.4.1b8ae20512db692f2a680001630.html

[5] **Experimental list of The Top 100 World Crises Growing with Growth** (Henshaw 2020):
https://www.synapse9.com/_r3ref/100CrisesTable.pdf

The systemic problem will eventually get the whole world's attention, of course. For now, though, responding to the real cause conflicts with popular belief systems worldwide. If our beliefs were logical, there could be a practical fix, but they are not. The realities we perceive are mostly given to us by our cultures. Our cultures then become divided into separate worlds that do not easily talk to each other. That leaves humanity without a common language for saving the Earth. Still, people worldwide are keenly interested in fostering successful new lives. Perhaps that could help foster the new common world language that seems to be gradually emerging.

## 2.2   When and How to Turn-forward

> *Strychalski says: "We still don't know the mechanism by which these things[biological cells] divide. That blows my mind— it is one of the basic aspects of life."* — Mitch Leslie (2021), *Science Magazine*

There is much we do not know about our surroundings. Still, we also often see danger ahead and know to turn away. We find it easy to turn away from danger steering a car, developing relationships, estimating budgets, or meeting deadlines. That may not work for dangers that only come around every few hundred years. What gets us to respond to risks is often gut-felt anxiety, a clear signal to face the future. We should be feeling that when thinking about the world's future.

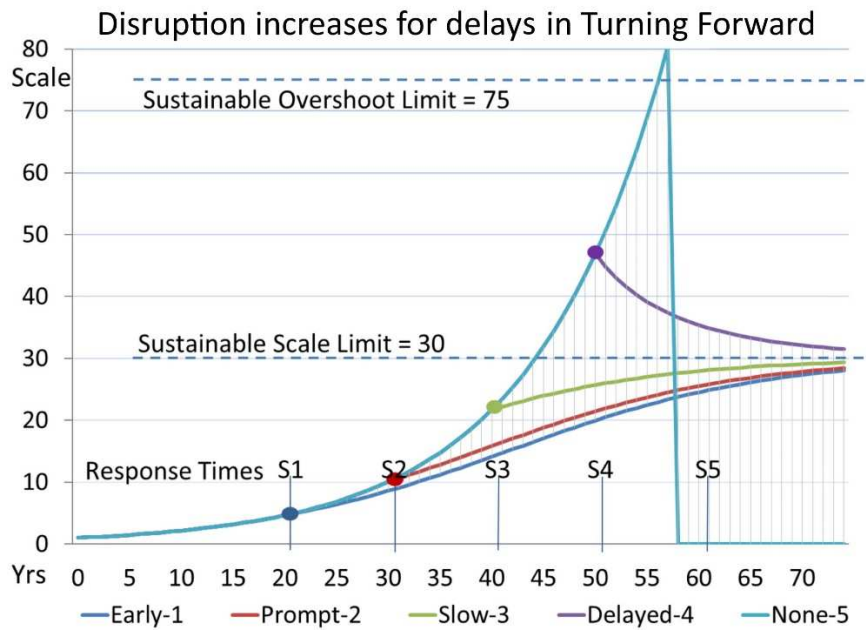

Figure 1.  Increasing Urgency for Delayed Turns Forward. Successive delays, S2, S3, and S4 result in increasingly disruptive responses to limits. The system has collapsed by S5 and so does not make the turn. (Henshaw 2010a)

Figure 1 shows a growth curve with a series of delays (S1 to S4) in starting to respond to approaching limits that should trigger increased anxiety. The scenario is fictional, and reality could be more complex. Still, the simple principle is that growing discontinuity and continuity conflict.  That would be the same independent of the units of time, scale, or limits. Early start times S1 and S2 cause no real disruption. Late starts time S3 and S4 do. Response start time S5 never comes. The main lesson is that early and late responses both end up around the same place at around the same time, so there is little sacrifice for responding early.

One potential loss from responding too early is the possible knowledge gained by dangerously overshooting and recovering. There is also a potential gain in system flexibility, in dangerously pushing the edge of chaos.

What Figure 1 does not show is the *point of timely response* when approaching threats are enough to motivate and not so late as to cause misjudgment. We experience that when skillfully driving, canoeing, sailing, making moves in politics, and other sensitive and intuitive steering arts. Responses need to gather first to make the right move and not be too soon or too late. Good steering waits for the right time, not just reacting as fast as possible. In reading Figure 1, imagine your anxiety rising on approaching the limit the way one feels making turns in one's favorite sport.

## 2.3   Steering the Living-Economy

*Nothing fails like success because we don't learn from it. We learn only from failure. — Kenneth E. Boulding (1978)*

The world economy has had a long success in repeatedly doubling our conversion of natural resources into wealth every 22 years or so (Figure 2). Judging from the atmospheric CO2 records (Henshaw 2019), rapid global economic growth appears to have started in earnest in about 1780 with the invention of Watt's rotary steam engine for powering boat and train travel and replacing waterwheel power for mills.[6] What is remarkable in these recent economic trends is that they each have 1) a constant growth rate and 2) remain in constant proportion to one another. That fixed mathematical *coupling* is clear evidence that each series reflects the world economy behaving as a whole. Moreover, they also all move together, indicating the whole economy has a stable driving positive feedback. Given the many forces driving economic growth, such as finance, government, consumers, institutions, etc., the pattern of growth constants says they all appear to work together.

All the data curves in Figure 2 (solid lines) closely fit the dotted lines for average exponential growth constants (Eqn 1). In general, that seems to be what is getting out of our control, the excessive amplifier feedback of income amplifying income, depleting the wealth of the Earth. Further confirmation that these patterns are whole system behaviors is in the convergence of the growth constant log plots.[7] But what is unifying the whole system's driving force?

---

[6] Watt steam engine https://en.wikipedia.org/wiki/Watt_steam_engine

[7] See §3.3 stat. test for homeostasis. Fig 15 http://synapse9.com/ISSS-21/NewSci-IndividSys-supl.pdf

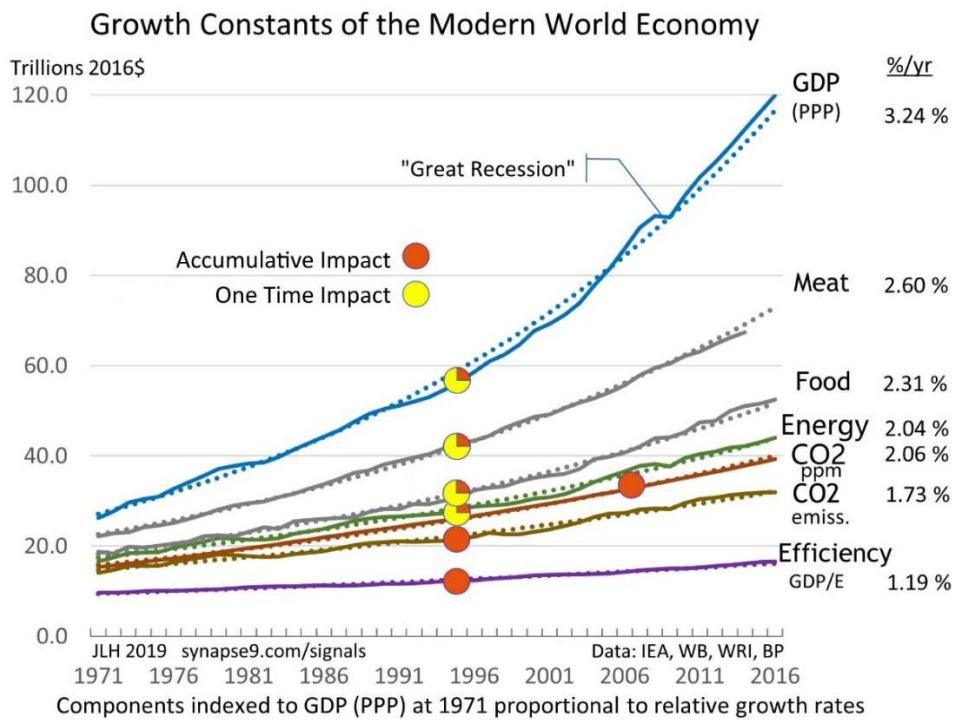

Figure 2. *Steady global economic growth rates*: Components are indexed to GDP (PPP) at 1971, scaled by the ratio of each one's growth rate to GDP's (%/yr). Data sources §8

Growth curve: $$Y = Y_0{}^r * t \qquad\qquad (1)$$

$Y_0$ – starting value, r – proportional rate of change, t – units of time

We seem to hear about that driving force every day in the news, everyone's desperate struggle to make the economy grow faster, to become more productive, as if necessary to escape from threatened poverty. There is the force of habit, too, of course. There are also many others, like the motivation of the scientists and entrepreneurs to create and invest in productive innovations. Individuals also invest in themselves to increase their productivity, also an important contributor. There is also a systemic feedback loop, the financial and business practice of taking profits from every source of income and using much of it to increase new investment. That sets up what one might call a *maximum carrot-and-stick* economic stimulus, both investing in businesses and escalating their competition, forcing everyone to increasingly struggle not to fall behind, a combination of virtuous and vicious feedback.

As the economy presses planetary limits, businesses still seem to benefit from the investment, but the profits go back to investors for increasing the competition. So, the pressure to overstep natural boundaries grows, destroying habitats and escalating everyone's struggle. The more the economy grows, the more intense the competition, not less. If we recognized the problem, it might cue the needed gut-wrenching response to "flatten the curve," as many learned to do for the pandemic.

Many global efforts aim to steer the world economy to safety, evident in the global sustainability movement. On might mark the start with the 1972 Limits to Growth study (Meadows et al.), largely confirmed in 2004 (Meadows, Randers & Meadows). The global waves of government and business sustainability efforts, including the UN's SDGs and the European degrowth movement *(Kallis et al. 2018)* and many others, growing urgency of economic system change. Today, many philanthropic investors are looking for environmental, social, and governance (ESG) impacts rather than only growing profits. However, the best-known sustainability efforts, corporate sustainability policies, the IPCC's climate change mitigation efforts, and the UN's SDGs all still seek to maximize growth to pay for correcting the harms caused by our pressing ever harder on the natural limits to growth.

Figure 3 shows a simple model that partitions the economy into two ways of making money, compounding profits in business and finance, circular exchange in commerce. Finance consists of banks, investor funds, and the finance operations of businesses that put money into *commerce* to finance the growing exploitation of nature and society, aiming to maximize the growing returns and reinvestment. Commercial operations also seek to maximize profits, but the great majority of revenue is spent, not compounded. Of course, the model is imperfect, as some profits for finance are spent without a promise of growing returns and some profits of commerce invested. We call *commerce* the freely circulating exchange of money for products or services without attached obligations for growing returns and finance the converse. The figure also loosely suggests the necessary financial solution; having finance discover the natural wisdom in spending (divesting) its profits to balance *finance* with *commerce*. It highlights the sticking point of an economy designed for exploitation to escalate.

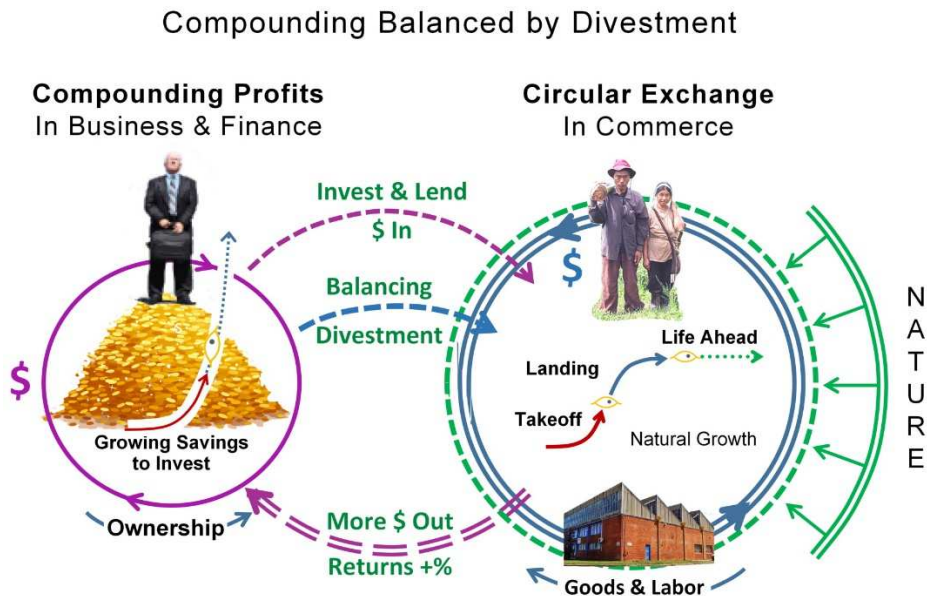

Figure 3. <u>Steering a Finance Guided Economy</u>: Business and Investor financial choices determine the use of the system's profits. Usually, it is to put money into commerce to take more out exponentially, called compound investing.

Just how the two come into balance matters greatly too. One might avoid an environmental collapse but not avoid a general financial collapse. Ironically a steady-state economy achieved by a steady divestment of profits could still be just as profitable as before, just like the family business that chooses to stay profitable and stop growing to take care of family.

If there's a broad consensus to relieve the relentless feedback, any sector might take the lead. It could be government, finance, business, nonprofits, social networks, thinkers, activists, or even liberals or conservatives, one at a time or all at once, like the broad consensus for sustainability, but focused on relieving the feedback and giving humanity a new life. In nature, the *turn-forward* from exponential growth to maturation can be disruptive if delayed or disorganized. So, we would want to find a smooth way forward. It will have to be more than one approach, it seems, considering the diversity of human cultures and deep sectarian divides.

————————————————

Spending capital to limit unsustainable growth is not entirely a new idea. In a conversation with Ken Boulding in 1983 about these general issues, he said repurposing the economy's profits to limit compounding as compounding starts to produce diminishing returns was one of JM Keynes' great ideas that did not catch on. In his General Theory (1935 ch 16 III, IV), he said (paraphrasing) that society might find better ways to spend the economy's profits as the worst of capitalism shows at its limits. Following that confirmation, it took years to tie together the many loose ends and find this general systems theory of new lives with a chance of being widely understood. Wide understanding seems to be the key requirement. Without a well-informed *planetary sense of community* (Francescato 2020), the world cannot effectively act as a whole and give itself a new life.

At least as important, Keynes also wrote about why people generally do not take risks based on their information. He saw that to take risks, people need to have information but then put it aside to act on intuition, what he called "animal spirits" (Keynes 1935 Ch12 VII; Akerlof & Shiller, 2010; Dow & Dow (2011). For society to take big risks, the strategy would be to study all the information and put it aside to act. Seeking a new life would be to set off on an uncharted course, and the world would need both great inspiration and a true desire to work together to make it work. It calls for collective motivation, a tidal wave of animal spirits, to see that making a new life for our world family is doing the right thing. Rebalancing *finance* and *commerce* would have other major benefits, too, such as funding greatly needed climate change relief. It would make the problems we face smaller and release needed finance (Keyßer & Lenzen 2021).

## 3   Growth Models for New Lives

### 3.1   Learning Nature's Creative Process

The cells of a human embryo multiply furiously during its first several months — multiplying to over a trillion cells at birth — to start the long maturation toward a vibrant adult life. We learn that sequence of changes in school, and it's part of our lives. This approach to studying it as a general pattern starts with noticing any significant change and then looking for growth processes by which it developed. That makes the change process just completed available for study. As one learns to spot the continuities to follow and milestone shifts in direction, one will also notice that the relationships tend to have environmental centers for different stages. Together they open up the study of the internal and external relationships that do the real work of emerging change.

An interesting example is how a house party or dance normally starts after some time after people arrive when a little contagious process seems to affect everyone at once to mark the real start of the party. Often there is also an after-party too. Then one of the favorite conversations is often the ups and downs of the party that preceded. That same kind of thinking is what one builds on to notice organizational changes of new lives in general.

Another familiar example of new lives to study is the process of "making dinner," either for oneself or a crowd. Preparation starts with an idea for an anchor recipe and shopping or taking food from the fridge and cabinets. After various smaller steps of getting organized, the plan usually solidifies, and the bigger preparation tasks get underway. Then when new ideas for the finishing steps arise and the effort is seen as a whole, it leads to taking out new ingredients and putting others away. At the turning point from starting things up to finishing them, attention shifts to side dishes and timing the end, adjusting the plan once again. Then others may pitch in as finishing touches are added, and cleanup starts before serving the meal. Then dinner is put on and takes a life of its own.

As we learn more about recognizing where and how new lives are developing, what may be most interesting is looking for their internal steering and external influences. For example, helping someone with learning challenges calls for a combined internal and external search for openings to the next level. It may take either a limited or extensive effort, trust-building, and rewards. That applies to the growth hurdles of personal relationships too. Numerous internal and external forces need to have their say. It applies to new businesses facing marketing challenges with start-up funds run low.

To study new lives, we also need a new scientific approach to using a fragile source of information, our recall of personal observations. We trust personal observations by default, but that is risky, especially when one's observations matter. So, it is important to learn how to make them reliable. We also need to rely on personal observation as a sole source of rich information about contextual relationships, or "warm data," as Nora Bateson calls it (2017). The scientific method is neutral about numbers but remarkably biased about complex living system relationships. For example, just having a new person in a group may raise questions about their perspectives and everyone else's. So, one change changes the whole. That is because it introduces new relationships, not just an increase in number.

Statistical modeling and abstract theory can still help raise good questions, even though representing nature with numerical models also strips away most of the evidence of contextual relationships one might like to investigate. Of course, personal observations often get cluttered with biases and suppositions, too, and social networks can reinforce the craziest kinds of ideas. So, learning to be an honest observer is itself a challenge. One basic method for learning to trust one's observations is to make a point of learning something new each time any issue comes up, not just triggering the same old answer over and over. There is always something more to learn.

Common language terms are another source of information on meaningful experiences and attached contextual relationships as *warm data*. To retrieve some of it, one can think through the varied usage of any term to search through its associations. Take words like "book" or "breeze." If you think of their uses, a rich variety of natural experiences and relationships comes to mind. Thinking about words that way helps one unpack the deep meanings of the cultural experiences we associate with our words and then the real meaning of terms for what is happening around us.

We also need the association of natural language with natural experiences for a special task. That is to help us distinguish things defined by nature from the perceptions of the world we define for ourselves. Just asking how what else you might learn about things helps with that. However, to use a potentially unscientific data source scientifically, we need a technique for testing our observations in exploratory ways to see what we can trust. There is more on that in §4.2

## 3.2   Partial and Full Growth Paths

*Every system is a "holon"--that is, it is both a whole in its own right, comprised of subsystems, and simultaneously an integral part of a larger system. Thus holons form "nested hierarchies," systems within systems, circuits within circuits, fields within fields. — Joanna Macy ( 2014)*

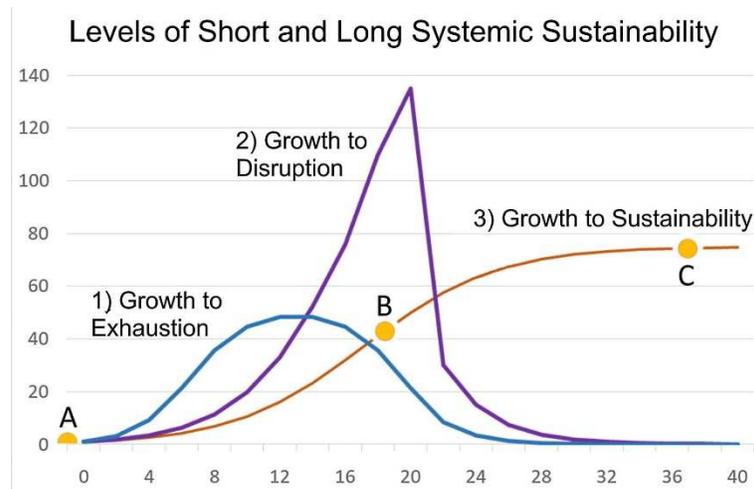

Figure 4.   *Levels of short and long-term systemic sustainability*: Systemic limits for maximizing growth: 1) consuming a limited resource, 2) hitting disruptive limits, 3) being responsive to both internal and environmental limits.,

The three simple growth curves in Figure 4 represent systems that fail or pass the tests of sustainable growth. All three curves start with a period of steady growth — curves 1 and 2 show new lives that fail to mature.

1.  *Growth to Exhaustion*: A growth system organized to multiply the consumption of its starting resource, but failing to find any other, such as — matches that flair but do not catch on — seed sprouts that fail to put down roots — a business that consumes its seed capital without attracting a market — enjoying a dish of ice cream.

2. *Growth to Disruption*: A growth system that finds more resources as they grow but overreaches internal or external relationships to become unstable, such as — seeds that sprout but become spindly and fall over — businesses that grow too fast and collapse in confusion[8] — diversified economies that die by killing their renewable resources.

3. *Growth to Sustainability*: A growth system that accesses new resources, responds to internal and external limits, develops resilience, and adapts to a secure niche, such as — plant and animal growth to maturity — the formation of communities and ecologies — self-limiting cultures, industries organizations, of professions — lasting family and personal relations — local businesses and nonprofits focused on serving their needs.

Read in this fashion; growth is an ascending scale of tests, familiar to all young lives, new businesses, and new personal relationships. To survive their growth, they need to pass the critical midpoint test of turning toward responding to the future, the turn-forward, a gateway to self-preservation often enough. Success and failure do not always follow simple curves, of course. It is only easier to learn from simple cases. Success may often come from surviving multiple failures, with every attempt a trial by fire. Then you might say it is following a strategy of *try, try again*.

## 3.3   The Principal Stages of New Lives

*"It always seems impossible until it is done." — Nelson Mandela*

What makes the growth of individual new lives an endless surprise is how they always come from such hidden and unique beginnings to then take on a busy world on their own terms. The series of events, environments, and stages of development diagrammed in Figure 5 is a map of what to look for and guide to the paths to follow and questions to ask along the way. The aim is to help people notice and respond to growth processes in their experience, past, present, and future.

It is a universal "hero's journey" of small beginnings and great challenges for individual systems with lives of their own. New lives also need to be opportunistic to survive and appear to have *minds of their own,* too (Henshaw 2008). That notable self-animation of growth systems also means having the ability to concentrate and control energy use, making growth a *syntropic* process in an *entropic* world, a natural pattern of the natural design hidden in plain sight. It is a universal pattern of how new lives develop that people already see and only need to connect with all the new lives around them and the state of our world.

Figure 5 shows the series milestones for new lives associated with the minimal "S" curve continuity for lasting systemic transformations. Natural growth systems might take more complicated paths, with detours, backsliding, recovery, and delays. The curve itself is "Nature's Integral," referring symbolically to how nature builds new systems by accumulating successive organizational changes. The shape of the curve is like that of a logistic equation. It is used here for raising questions about punctuated organizational changes of natural growth. Nature is not predetermined and needs adaptive building processes to determine how systems will develop.

---

[8] Eight Dangers of Growing Your Business Too Fast https://www.inc.com/cox-business/eight-dangers-of-growing-your-business-too-fast.html

## The Growth Milestones of Individual New Lives

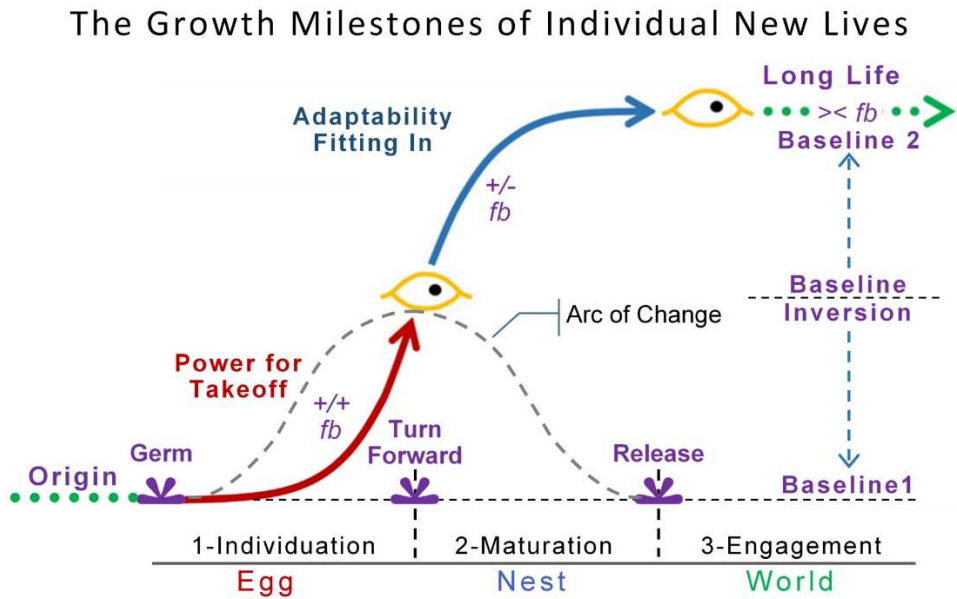

Figure 5. *Nature's Integral (the curve)*: With three events (Germ, Turn-Forward & Release) initiating three directions of growth (Individuation, Maturation & Engagement) in three successive environments (Egg, Nest, World). The 'eyes' between periods indicate likely anticipatory changes in direction.

The complex series of changes highlights the complications nature needs to create a smooth organizational development process. The red curve represents what occurs in the protected confines of *the egg*, or womb, garage, or other quiet places where new life can germinate and multiply to gain power during its takeoff period of *individuation*. That big leap of positive feedback relative to the origin needs to give new lives the ability to fend for themselves in the *nest* environment after the *turn-forward*. Then as their physical *maturation* begins in the *nest* following the blue curve, the growth shifts to negative feedback for approaching the future, going slower and slower, perfecting the fit. That period of exposure to the new *world* they will enter prepares the new life prepares for the next big leap of *release* for *engagement* in that world. The dotted green lines at the beginning and end represent the busy *worlds* that new lives emerge from and into, open environments where complex niches and new forms of relationships develop

Let us look more closely at the most hazardous transition, the turn-forward when the new life moves from its *egg* environment to its *nest* environment. At that risky midpoint of development, there is a somewhat abrupt switch from backward-facing to forward-facing development, from multiplying a seed pattern to adapting to a new world ahead. One might ask why that switch happens? It appears that it is what evolution discovered for sustaining life, a whole system test of adaptivity. A seed pattern cannot possibly encode a whole future environment that will challenge a new life. So it appears life had to become exploratory and actively adapting to unfamiliar environments as the future arrives. It makes life possible.

That shift from referencing the past to referencing the future is also clear in the shapes of the red and blue curves. Each step along the red curve is a change proportional to the total change in the past (referencing baseline B1). Each step along the blue curve is a change proportional to the distance from the destination (referencing baseline B2). That change in baselines for growth can also repeat. Everything we struggle to learn can be the foundation for learning more, a series of growth processes. A familiar example is the annual succession of grades in school, each year a separate growth process, chained together.

In Figure 5, the dotted grey curve in the middle, labeled *Arc of Change,* shows a rising and then falling pair of "S" curves, as the arc of the physical development process, having four growth curves, each with beginning and ending processes. It offers a new perspective on the complicated underlying rates and accelerations of change. If you have good data for a growth process, that arc of change is a plot of the slopes of the growth curve, or first derivative, and exposes the growth process as even more of an assembly line of processes linking together.

## 3.4   Basic Study Tips

A study of the growth of new lives can start with making a list of the many kinds of new lives you are already familiar with, including the growth of one's own life, career, endeavors, relations. Then add others one finds interesting. Next, using Figure 5 as a model, practice with pen and paper and post-it notes, drawing the intuitive shapes of growth stages for some of the items on the list. Keep it simple at first, and maybe put it aside to think over. Subjects might include the arc of preparations for a memorable dinner, events over a summer season or school year, or the history of a business or social network campaign. If data is available, use that as the *primary indicator*. If not, what works best is to pick a primary indicator intuitively and try some *secondary indicators*. Then label the development milestones you see, the *germ*, *individuation*, *turn-forward*, *maturation*, *release,* and *engagement* milestones, that shape the arc of the whole story.

The next step is to redraw the curves on the same timeline, letting the drawing get a little messy until they most accurately reflect the rise and fall of the main developments and turning point events, as the arc of their story (Henshaw 2018, 83-87). After that, put the work aside for some hours or days, and upon returning, redraw and relabel a final version of the curves. Then edit your notes on the *individuation*, *maturation,* and *engagement* periods, on the influences of their *egg, nest,* and *world* environments, and on what you learned. One might then move on to other subjects, events in history, familiar social movements, school or work career, or current problems like growth hurdles of children or organization projects. It gets one thinking in a way to notice more details about surrounding changes. When time-series data is available, how the data points depart from the trends might suggest things to investigate. Data that departs from trends could be noise, of course, but that would *not* be assumed. Departures are quite often evidence of other signals, possibly evidence of systems on other timeframes or scales, important to discover.

For example, a break in conversation might signal a distracted partner or possibly something more that needs to be said. In either case, it might well not be random. In this way, various kinds of inquiry can help raise useful questions and expose subjects to investigate. The raw data for complex flows is often mistaken as noise when it is actually produced by complicated fluctuation. For that, a combination of careful statistical tests for implied continuity and careful methods of derivative reconstruction to partly recover it are needed (Henshaw 1995, 1999, 2007). That gives raw time-series data a special role in investigating natural systems because it captures traces of the complex continuities of the physical processes it reflects, a type of *warm data* hidden in *cold data*, exposing deeper relationships.

Time-series data lets the continuities of nature speak for themselves. Several case studies for working with data are in the Supplemental Topics.[9] One might get data to study from the accounts manager at the office to plot the hours spent on office projects or campaigns. Towns and cities have time-series data to start researching their growth too. When those are plotted over time, with markers for various milestones, see if the growth trends and inflection points show the liveliness of a living system. Published environmental and economic data is a good source to study too.[10] Unfortunately, stock market data does not work well, so influenced by rumors and tricks; it rarely reflects continuities.

## 3.5   Related Transformation Models

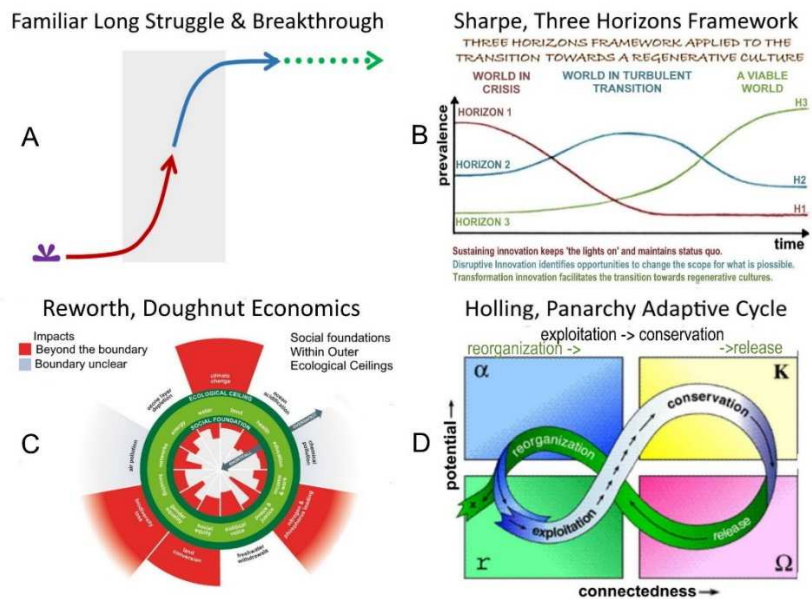

Figure 6.  *Four popular related system transformation models*: A) familiar "Long Struggle & Breakthrough," B) "Three Horizons" Sharpe (2016), C) "Doughnut Economics" Reworth (2004), D) "Panarchy Adaptive Cycle" Holling & Gunderson (2001, 2002).

---

[9] Supplemental Topics - Case Studies §2: http://synapse9.com/ISSS-21/NewSci-IndividSys-supl.pdf

[10] Data transformation: Google Trends https://trends.google.com/trends/?, Google Word Use History - Ngrams https://tinyurl.com/myxbpwh5, NY Times https://www.nytimes.com/, The world Bank https://data.worldbank.org/, Our World in Data – https://ourworldindata.org/, WorldData.AI https://worlddata.ai/

It is valuable to learn from several transformation models, combining complementary strengths. The four popular models of whole system transformation in Figure 6 are all related to the general model of *individual new lives* in Figure 5. Any of them could be one's primary reference depending on circumstance, with the others referred to for enrichment and finding what pieces fit best. Together they help expand the tool kit of principles and intuitions about whole system transformations. Over time, as people gain experience and others study the subject, various perceptions will combine, and community standards develop. Referring to Figure 6:

A. *Long Struggle & Breakthrough* – Everyone is familiar with repeated failures that lead to success, shown here as a compressed version of Figure 5. Long-delayed breakthroughs might come from delayed maturation, the appearance of new openings, or dramatic leaps. One way or another, the environment eventually opens to the efforts before they are exhausted. A good example is the delayed breakout of hip-hop as a national culture after 20 years of circulation in local New York neighborhoods. Then, in the early 90s, it met sudden economic success coincident with New York youth cultures turning away from previously dominant drug cultures[11].

B. *Three Horizons Framework,* Sharpe (2016) – The strength of the Three Horizons Framework is its ability to coordinate three groups to work on different phases of a long-term strategy. One group can work on breaking away from outmoded practices (H1). Another can work on transitional stages (H2). Finally, a third group can work on the long future (H3) as they all coordinate. It is like fast-tracked building designs with demolition and design starting together. Parenting also works on short, medium, and long-term horizons. Business planning and community organization do too. Long-term world economic plans have followed multi-phase models, too, like the rebuilding from WWII.

C. *Doughnut Model of Sustainable Economies*, Raworth (2004) – The doughnut model combines thinking on internal services and external restraints so that urban plans can optimize community benefits within planetary boundaries. The limitation is that the nine planetary boundaries of growth (Rockstrom et al., 2009) are far from inclusive[12] and mostly are not traceable to individual business or municipal choices. So, more collaboration is needed to distribute responsibility fairly for *all* GDP impacts so that global thresholds can be respected (Henshaw, King & Zarnikau, 2011; Baue 2019).

D. *Panarchy Adaptive Cycle*, Holling and Gunderson (2001, 2002) – Panarchy is a cyclic model of environmental evolution, similar to incremental seasonal cycles. It also partly fits how economies evolve, with cycles of growth followed by climax, recession, and reorganization. The terms of Panarchy start with new growth (called *exploitation)* followed by maturation (called *conservation* — leading to stasis), then system breakdown (called *release), and reorganization,* after which the cycle repeats. For ecologies, cyclic regeneration is not recovery from a disordering collapse, though. So, one reads the model metaphorically. The phases of Panarchy roughly correspond to the *egg*, *nest* & *world* for *individual new*

---

*lives* if one adds phases of decay and rebirth. Those phases were left out of the *new lives* model to focus only on the details of how growth can lead to long lives.

Left out here are also a great many other models for system transformation. Salk & Salk's "A New Reality" (2018) offers a very compatible 'S' curve development model to turn the reader's attention to approaching global end-of-growth transformations. It is particularly easy to understand and suitable as a children's book. Duane Elgin's *Choosing Earth* (2020) is another helpful view of changes to future society in the Anthropocene, emphasizing a well-researched but more stark view of likely societal crises ahead. The book takes the reader on an extraordinary journey, extensively documented with authoritative references, for appreciating the large changes to living on Earth our children might likely face.

The transformation models of leading world institutions, like the UN's IPCC and SDG efforts, and world governments unfortunately still project a regular doubling of the economy's demands on the Earth and society, ignoring the role of growth in nature for creating new lives, a strictly one-way and relatively short-lived process. Yet, even steering the world looking in the rearview mirror, as we are, many new directions can already be seen emerging.

Complexity science is another highly funded approach for studying complex systems and their behaviors. Unfortunately, its experiments with complex systems models have been economically important but less successful for interpreting real economies in real environments. Worth mentioning, though, are two examples of often forward-thinking complexity science, the consulting work of Yaneer Bar-Yam (2004) and David Snowden (2021). They associate the statistical properties of complex systems with informed decision-making, creating practical search strategies for real-world applications.

## 3.6   Systemic Strategy

*I may briefly remind the reader how little we can trust to our unassisted senses in*
*estimating the degree or magnitude of any phenomenon.*
*— Stanley Jevons (1887 p 276)*

Developing a successful way of intervening in or managing living systems depends on reliably observing available openings. Every stage of development is an opening in some directions and a closure to others. That enables some kinds of efforts and inhibits others. Because conditions frequently change, one needs always to be ready to rethink and look for new openings. Because of the human attraction to magical fixations, people often commit to misguided efforts, too, so a way of carefully vetting commitments is needed. One way is to impersonate the system in question, asking: "What a building wants to be?" as a way to search for holistic insight. Louis Kahn, the marvelous 20[th]-century architect, was famous for seeking out the intentions of a building during design, frequently asking: "What does it want to be?" In one case, that led him to discover that the inner sanctuary of a great new mosque wanted to be the building's main entry (Kahn 1998, p 44), an unexpected method for designing places that come alive.

Elinor Ostrom (1990, 2009) and Donella Meadows (1999) have provided quite helpful lists of guiding principles for intervening in (Meadows, Table 2) or managing (Ostrom, Table 1) complex living systems. Notice how Ostrom's principles have more of an external view of the subject. Meadows' principles are more for working with living systems from the inside. Some of the author's related systems steering principles are in Table 3.

Table 1. Twelve Leverage Points for Intervening in Systems (Meadows 1999):

1) The power to transcend paradigms.

2) The mindset or paradigm from which the system, its goals, structure, rules, delays, and parameters arise.

3) The goals of the system.

4) The power to add, change, evolve, or self-organize system structure.

5) Information flows.

6) Material stocks, flows, and nodes of intersection.

7) The driving positive feedback loops.

8) The strength of negative feedback loops relative to the forces to mitigate.

9) The lengths of delays relative to the rate of system changes.

10) The sizes of buffers and other stabilizing stocks relative to their flows.

11) Constants, parameters, numbers (subsidies, taxes, standards).

12) The power of small nudges

Table 2. Eight Principles for Managing a Commons Ostrom: (1990):

1) Clearly defined boundaries.

2) Graduated sanctions for respectful compliance.

3) Govern the use of commons goods according to local needs and conditions.

4) Allow those affected by the rules to participate in writing the rules.

5) Make those who monitor the rules accountable to those regulated.

6) Low cost and accessible dispute resolution.

7) Right to self-organize regulation systems.

8) Organize governance of nested boundaries in layers under the whole.

Table 3. Eleven General Principles for Self-Organizing Transformation (Henshaw)

1) Local solutions also need to support larger system changes.

2) Staying in touch with diverse channels of communication.

3) Noticing emerging new lives opening channels.

4) Building the common human language while respecting individual cultures.

7) Noticing the openings for change, each stage of growth a different one.

8) Noticing there may be no technical fix for economic transformation, but it will also take great technical sophistication.

9) Noticing how productivity comes from connecting complementary parts.

10) Noticing where diversity thrives, like in ponds, forests, towns, and cities.

5) See systems as having lives of their own, having wisdom, and needing guidance.

6) Noticing that humans, like animals, spend their time searching for new paths.

11) Noticing how self-organizing systems distribute the pressures put upon any part, like rising competition for productivity.

## 4   Roots in the Sciences

> *"The impossibility of a truly comprehensive account of growth in nature and society should not be an excuse for the paucity of broader inquiries into the modalities of growth" — Vaclav Smil (2019, p XIX).*

The origin of studying growth as nature's formative process seems to go back to proto-Greek nature-science. The common Greek word, Φύσς, pronounced *phúsis,*[13],[14] refers to nature's productivity, growth, and birth. However, when the same term for nature was adopted by Greek scientists, it came to refer to abstract conceptual properties and theories for nature (Aristotle ed. 1941, Bäck 2006). The term was then later used for the modern discipline of physics. History is difficult to read, but the two opposite meanings do exist, and nature is both creative and determinate. It is also clear that the abstract deterministic theory meaning is what later became dominant. There could well be more to the story, but the Greek language is much older than Greek science. Greek science is also often referred to as being a rebellion against what the scientists called *nature-religion*[15]. Perhaps *nature-religion* was just a pre-abstract form of science that abstract science chose to erase and perhaps should not have. In any case, that interpretation is reassuring for the principled *post-abstract* form of science offered here.

We are, of course, also greatly indebted to modern science for all the rules of nature we can know with certainty. Moreover, the scientific revolution and method offer a wonderful discipline for converting data to useful equations for guiding innovation. Still, representing nature as following our abstract rules limits the questions one can ask, blinding us to living systems we rely on all around us and to nature's purpose in starting all new lives with compound growth. It is not primarily to give humans extraordinary power over nature and each other. It's to give new lives a kick-start, a dramatic change of scale giving them a fighting chance, one we should use.

There have been many scientists who also bucked the trends, more than one can name. Some of the notable scientists who challenged precedent and contributed to the development of this diagnostic view of natural systems include: Johann Goethe (ed. 1996), D'Arcy Thompson (1942), Ken Boulding (1953), Garret Hardin (1968), Gregory Bateson (1972), James Miller (1973a, 1973b), Maturana, Varela, & Uribe (1974), Brian Goodwin (1982), Elenor Ostrum (1990), Robert Costanza (1997, 2014), Donella Meadows (1972, 2001, 2004), and Vaclav Smil (2019). Brief discussions of the contributions of Boulding, Hardin, Smil, Ostrom, and Meadows are below. Discussion of other contributions that helped lay the groundwork for this study is in the Supplemental Topics.[16]

---

[13] Wiktionary: Φύσς) Translated "gro.sis" and pronounced "fi.sis." https://en.wiktionary.org/wiki/%CF%86%CF%8D%CF%83%CE%B9%CF%82

[14] – Merriam Webster: physics; History and Etymology https://www.merriam-webster.com/dictionary/physics

[15] History of physics https://en.wikipedia.org/wiki/History_of_physics

Vaclav Smil's (2019) study of a wide variety of growth systems is related to the present study as a diagnostic study of growth systems aided by simplified models. His approach is to construct mathematical models for the shapes of all kinds of environmental, societal, and economic growth systems. Smil does not plot raw data with the growth curves discussed, though. Showing raw data overlaid with some of the model outputs might a more lively appearance to the metabolic processes of lively systems. Still, the growth curve model plots do help illustrate the carefully researched metabolic processes he discusses.

Ken Boulding's "Toward a general theory of growth" (1953) is a general theory of growth as a pervasive natural phenomenon. He was an economics student of Keynes, and perhaps they influenced each other. Both seemed to study growth as a process of natural system development from a diagnostic rather than an abstract theory approach. The paper identifies several general metabolic principles, paraphrased here for clarity:

1) *Nucleation principle*: Growth requires a nucleus to initiate growth in most cases: a) the condensation of raindrops needs a dust spec to start, b) the retention of information on new subjects requires some new insight to nucleate in student's minds, c) the student ability to improvise greatly improves whey they realize that language has grammar.

2) *Nonproportional scales principle*: Changes of different dimensions of a whole tend to have different scales, such as length, surface area, weight, volume, temperature, appeal, etc.

3) *D'Arcy Thompson principle*: The proportions of natural systems come from the relative extents of their growth, i.e., growth defines form.

4) *The Carpenter Principle*: Growth exhibits unexplained coordination of the whole as if a carpenter makes all the parts fit, requiring the parts to mutually adapt, as if to a common plan.

5) *Principle of Equal Advantage*: The parts of systems fit together to maximize their potential (as if for economic advantage), a corollary to the carpenter principle.

6) *Principle of Natural Pace*: The coherence and cohesion of a growth process may break if growth proceeds too long or ends too soon, if growth rates are too high or low, potentially even causing collapse. (fragility rule)

Garrett Hardin (1968) explored the class of systemic problems with no technical solution. His first example is of an arms race, in which adversaries increase their military power but decrease their security. His most famous example is of an economic crisis caused by blind self-interest in a small village. The villagers share a meadow for grazing milk cows – its commons – which becomes overgrazed. At some point, someone pastures one too many cows, and the productivity of the commons starts to decline. To keep getting enough milk, each family increases the number of cows it puts out to pasture. That tragic strategy of growing investment in a depleting resource spirals out of control, leading to the meadow and the cows both becoming barren. The "tragedy" is then really of the villagers' fixation on a failing strategy. They do not see the harm until it is too late, if even then. The problem is conceptually simple but mentally complex. All living systems that outlive their growth begin with compound growth and turn forward to adapt to natural limits, but Hardin's villagers cannot, as if subject to fictional realities.

---

[16] See also §4 Roots in the Sciences: http://synapse9.com/ISSS-21/NewSci-IndividSys -Supl.pdf

That dilemma closely matches our global fixation on compound growth in income obtained by multiplying our demands on the Earth and societies, like Hardin's villagers, mistaking ever-rising income for increasing wealth, not noticing it has become the opposite. When it changed seems likely to have been some 60 years ago, when the environmental crisis first became big news. Societally we placed a higher value on maximizing growth than preserving our natural assets. As a result, our endless growth policy and the fabulous riches it seems to create became a very dangerous Faustian bargain, degrading our primary asset like an overgrazed meadow.

So, we have a problem with no technical solution. To solve our tragedy of the commons, we likely need the details of the problem to become common knowledge so that no one can hide from it. There indeed is a growing worldwide feeling of unease about our growing swarms of serious environmental and societal challenges. However, if it will blossom and shine the glare of daylight on the problem is hard to say.

## 4.1   Origin of Natural Systems Science

This work started in the late 1960s with a series of college physics labs studying why every run of any experiment somewhat misbehaves and curiosity about why natural shapes have rounded corners and edges. Discussions with grad school friends in a Brooklyn apartment brought out the puzzle of *individual differences*. Those interests bore fruit a decade later, after studying architectural design when doing field research on the microclimates of homes (Henshaw 1978). Each home study combined 24-hour chart recordings of temperature and airflow with smoke tracing for individual air currents and current networks to map the indoor climate. In each home, airflow networks repeatedly formed and were replaced throughout the day as the sun moved from east to west and then faded throughout the night. Sometimes a column of warm air would rise from a warm surface and accelerate as it grew until it broke its connection with its source, creating a free-rising individual vortex. That led to the insight that growth systems inevitably change conditions until they interrupt their designs and turn into something else (Henshaw 1979, 1985a 1985b).

In the 1980s and 90s, research turned to algorithms for finding and diagnosing continuity in time-series data, an apparently novel approach. It was a chance to experiment with ways of decoding creative natural processes reflected in time-series data. The algorithms developed for the diagnostic study of complex ecological, astronomical, social, and economic systems, called *derivative reconstruction,* exposed many hidden features, like inventing a telescope, opening uncharted territory (Henshaw 1995, 1999, 2007). General theory papers followed, focusing increasingly on the active learning exhibited by many kinds of natural and human systems (Henshaw 2008, 2010a, 2010b, 2011, 2015, 2018, 2019). Over time the shapes of the data curves started being studied for the implied internal organization of environmental systems. Networking and work in UN and civil society organizations broadened the perspective.

The principal model for natural growth presented here, *nature's integral* (Figure 5), originated from many years of observing the same beginning and ending dynamics, then tracing its origin to an obscure implication of energy conservation. The principle that energy is neither created nor destroyed prohibits infinite rates of energy transmission, requiring nature to use developmental processes, the simplest of which is the "S" curve (Henshaw 2010b). The study of how alternating organizational development processes produced the shape then exposed the universal series of growth milestones that shape the development paths of new lives. Of course, more complicated behavior is also common, but it seems to be the smoothly flowing processes that exhibit the highest degrees of internal organization.

# 5   Background on Individual New Lives

## 5.1   Natural Systems Theory

In summary, *individual new lives* are systems that develop energy concentrating (syntropic) organization, short or long-lived, growing from an individual germ, seed pattern, or nucleus to develop the internal and external systems of relationships. They are also uniquely individual due to unique starting patterns and circumstances developing individually. Later, their growth takes them into new environments, where they also need to adapt.

There are also energy distributing (entropic) systems. One might call them i*ndividual new waves*, for symmetry, from the energy thrown off by the syntropic processes of individual new lives. For example, a person might throw a pebble in a pond, creating entropic waves after concentrating the energy to throw it. Generally, the organization that allows *energy concentrati*on is temporary, and energy dissipation follows. That alternation of energy concentrating and dissipating processes of new lives appears to be what makes life lively. The terms alive and animate would be reserved for biological life. It is important to notice that organization in nature heavily relies on the continuity of intermittent relationships. It sounds contradictory but is as simple as relationships based on intermittent contact and evident throughout ecologies, societies, communities, businesses, etc. So, one might draw almost any natural continuity as a dotted line for any natural network.

Many ecologies and societies are old enough to have uncertain origins, too, so though they may have emerged by growth, now they evolve in both drifting and growth-connected ways as new cultures within them emerge, change and fade. Nevertheless, they also appear to be positively energized and have sustained individuality over time. They also support the growth of diverse sub-cultures such as the social, political, artistic, and other human cultural movements we recognize. So, the *individual new lives* model needs to expand to a *new lives host* model for ecosystems. One might also use Miller's (1973a & 1973b) or Varela, Maturana, and Uribe's (1974) ways of defining living system boundaries or refer to their centralized nests of feedback loops. Also useful are functional features like neighborhood boundaries, a common language, or having the same employer (Henshaw, King & Zarnikau, 2011).

Individual people can also have active roles in multiple communities at the same time. This fluidity of cross-connection is an important natural property seemingly responsible for why ecologies and economies can thrive, referring to as a *semilattice* design in Alexander's "A city is not a tree" (1965). Such loosely bound communities, cultures, and economies also very interestingly can retain their identities through intermittent continuities without being contiguous. Having self-contained feedback loops seems to be the common thread. That makes them hard to follow and only a little easier when following the dynamically bundled feedbacks of growth as a tracer.

Perhaps the most important distinction for new lives is whether they mature and remain lively after they end their growth. That is a common property of successful biological lives, their communities, cultures, ecosystems. Individual businesses and institutions achieve thriving stable states for long periods too, such as family businesses, business centers, schools, and other local institutions that anchor many communities. One of the good indicators of *new lives* reaching that elevated stable state is *exploratory behavior*, such as new lives display as they move from their nest to world environments.

## 5.2   Recognizing new lives in formation

> *"There is a rare property of mind which consists of penetrating the disguise of variety and seizing on the common elements of sameness."*
> *— Stanley Jevons (1877)*

To get comfortable with this approach, spend some time looking for growth and change in subjects one finds fascinating. One can even watch the development of thoughts and feelings that develop slowly enough. For example, as we try to quiet our thoughts, we may also notice new thoughts blossoming from feint intuitions and become full-blown. Getting accustomed to thoughts emerging by a growth process can help with retracing the nascent thoughts they came from, took root, or drew reactions.

An office environment is another good place to watch things grow. Watch how new office plans or conversations come together, noting contributions from multiple sources. If things develop too fast and overshoot their limits, try to notice how they fall apart and what parts are leftover. Try to find the germinating center and see if the end preserves the original character. Many offices have established methods for growing their projects. Compare them with this method and classical methods like Action Research and Pattern Language (Henshaw 2018).

This approach to understanding living systems relies on a combination of honest personal observation and holding onto raw impressions. Journaling is a traceable way to collect raw data while looking for things to study in patterns of growth and change. Signs of things worth looking into include troubling or curious discoveries, unexpected environmental or social change, absence or calm, places where things often develop. So, look around to see if things happening come from some quiet place nearby. Once one finds a hint of something happening, think of it as a cell of relationships and mentally trace events as far as possible. If some threads have a history of development, look for signs of momentum and other connections. If it is something new and might be on a path of greater change, ask if it is something to avoid, oppose, just watch, or befriend.

Growth processes in nature begin with multiplying from small beginnings for another reason too. If a single cell in the womb grew at one cell per second rather than doubling every week, pregnancy would need to last 900 years, not nine months. That *egg* phase of growth is a radical way of speeding things up. It then takes another dramatic non-linear process to slow it down again, halving the rate of change again and again in the *nest* phase. That is what maturation does, slow down the accumulation of changes to stabilize the whole, perfecting things rather than multiplying them.

## 5.3   Using Natural language

Natural language is a particularly rich source of meanings people have found for natural experiences throughout the ages, making it a great source of information on relationships. Of course, it is not always neutral but still an excellent source of warm data on many subjects. Common old terms like "journey" and "home" tend to have especially rich associations. What helps bring them out is thinking of all the many variations of their use. Words that are easy to connect with root experiences are common words like *cold*, *stick*, *fruit*, *wave*. Other terms like *confusion*, *congestion*, *practice*, and *anticipation* also raise emotional reactions linked to common experience, exposing the taproot meanings if you look for them. Reconnecting with those root associations can clarify both basic issues addressed and help identify cultural meanings that have been added (Henshaw 2015).

Fortunately, most natural language terms were coined to recall important natural events and experiences and retain those root associations. It makes language a rather faithful recording of what people found worth referring to, a kind of map of life one might learn to read. Because growth processes seem self-similar on every time and spatial scale, the meanings of familiar ones might expose meanings that apply to others, too, like how noticing that air currents develop by growth led to this general study.

## 5.4   Distinguishing Terms

*"Remember, always, that everything you know, and everything everyone knows, is only a model." — Donella Meadows (2001)*

*Though statistical data is useful, it is also limited due to the common practice of decontextualizing the focus of inquiry.*
*— Nora Bateson (2017)*

The method used here for validating observations involves shifting attention back and forth between new and old impressions, not to affirm our interpretations so much as to enrich them. It is something like an archeological view, studying a fossil to discover new details each time to become part of the interpretation, as continued exploration. As a scientific method of observation, it helps distinguish what nature defines from our interpretations. We need to store up our interpretations of things, but we also need to make a habit of always learning something new. It essentially treats interpretations as questions and relies on preserving raw impressions from which to learn more.

To be honest with ourselves, new observations need to produce new information. So, to validate observations, we need to be open to the unexpected and explore new approaches rather than check things only to confirm what one already thinks. It does take practice, though. It means actively approaching observation in new ways and making a point of learning more each time to avoid simply reinforcing old interpretations. This openness to what else is there sharpens the mind and keeps it honest. It is not always simple, of course, as we all have unexpected assumptions. Trying to see more of what is there does bring improvement, though.

Terms referring to things defined by nature may often need clarification. For example, anything given a name immediately begins to accumulate attached meanings, and mixing up the things with their meanings is easy. In his living systems theory, Miller (1973a, 1973b) also maintained a distinction between naturally and mentally defined subjects, referring to them as *concrete* or *conceptual*. There is also terminology confusion for words that have different meanings in different cultural contexts. Language is so rich it lets us refer to all three, the cultural, abstract, and natural definitions, all at once, without realizing it.

For a simple example, the meaning of a phrase like *the box* depends on the context. Without clarification, it can refer to a particular container, a physical or mental trap, a sporting formation, a shape drawn on paper, or perhaps the strictures of a new business or personal relationship. Each of these interpretations attaches a cultural meaning to things defined by nature. Distinguishing between them may be hard to do without practice. Artists and archeologists, as well as detectives, parents, and scientists, often have practiced methods for learning from exploring what nature has defined, so they make discoveries such as finding that niches are safe zones with many visitors.

## 6   Discussion

Like most things, the development of modern science rested on its economic success. It gave us amazingly useful knowledge and three centuries of dramatic economic growth. But, unfortunately, science also represented nature with numerical models, stripping away crucial contextual relations and blinding us to the organizational designs of nature. That enabled the economy's ever-growing destructive exploitation of people and the Earth. Proposed here is the beginning of an observation and data-driven method of studying the designs of natural systems, one that might be widely understood and used. There has been both repetition and things left out, of course. That is partly because the primary aim has been to give readers a general picture and practical approach to building upon their own insights into the *new lives* we are part of and surround by; how they develop and take on roles in the world. Of course, the *new life* of humanity is also both of very special concern and a source of hope.

The current progress with the subject came from many years of fascination with it. So, following one's fascinations might be a good guide to success using it, balanced with other techniques. Leaders of every kind might also find success by studying the fascinating new lives throughout their worlds to help them better understand what to do. Only real understanding will persuade the great diverse cultures of the Earth to turn their world views in new directions as our relation to the earth changes.

Also left to the reader are many details of exactly how to rebalance the two main ways of making money termed *finance* and *commerce* in Figure 3. As long as the growth milestones of new lives remain hidden in sight, it will remain hard to explain. However, as people begin to see the pattern, plausible responses will be easier. An old model for how we could design institutions to guide us, for those interested, is a 2014 proposal for the UN SDGs, called The World SDG.[17] In short, it proposed collaborative economic and scientific research first to produce ratings for sustainable investments, followed by societal approval, to give investors sound choices and leave societies in control. Perhaps somewhat connected is a related approach, led by UNEP-FI, for guiding the global response to climate change.[18] Our global shift from maximum carrot-and-stick growth to natural growth would need affirmative guidance of that kind too.

People worldwide care deeply about their own lives, the *new lives* of close relations and communities, and the planet. The hope is that the general pattern of growth milestones, which everyone already knows intuitively from life experience as a simple fact of nature, will rise through the clouds of discord and uncertainty to allow us to act together as a world community. Many people of all kinds seem to agree already that our world needs a whole new life. Another positive sign is the apparent global spread of grassroots culture-change communities that might help spread the idea of following the natural growth patterns. Donella Meadows' Dancing with Nature (Meadows 2001) also conveys the idea, suggesting ways we can concretely move away from our war with nature into dancing with our ultimate home and family.

What might add an important extra level of confidence in this approach is that this understanding of new lives rests squarely on one of the cornerstone principles of natural science, that energy is absolutely conserved and can neither be created nor destroyed. Once that principle is associated with the implied continuity in organizational processes (Henshaw 2010b), it might stimulate a wide variety of new kinds of natural science research. For example, one related field of pattern recognition needing work is on exposing hidden continuities in irregular time-series data, expanding on the method called *derivative*

---

[17] The World SDG, https://synapse9.com/signals/2014/02/03/a-world-sdg/ See also
https://synapse9.com/signals/category/policy/ for other old public policy journal entries

[18] UNEP-FI Alliance *https://www.unepfi.org/net-zero-alliance/ and /net-zero-alliance/resources*

*reconstruction*[19] (Henshaw 1995, 1999). In general, there is a great need for a better understanding of the future too. Curiously, finding that natural change requires continuity could also contradict how quantum theory describes change as occurring without duration, another cornerstone principle of physics. The explanation might be simple, though, maybe only saying that quantum theory is a statistical theory rather than a physical process theory. That leaves both theories just as useful but independent. Hopefully, someone reading this will explain the discrepancy more fully.

That nature's purpose in starting all new lives with compound growth is to give them a kick-start leap in scale, leaving them fully-formed and ready to develop lives on their own. That is the heart of what has been hidden from view. Now we can see it as clearly visible throughout history and nature. The future has always come from a kick-start followed by a resolution of the growth of new lives. With a better understanding of that purpose of growth and the importance of a timely *turn-forward*, we can do much better than we have, and where appropriate, making growth our reliable servant and partner, not our master.

As for many new lives, the *turn forward* upon leaving our *egg* environment of boundless growth may come as a huge shock to the system, landing us somewhat incongruously in the protective *nest* of our new perception of nature. Many of our favorite old habits of boundless growth in the *egg* are sure to be left in some disarray. Shock can also be protective, too; let us hope. The dangers we might face in the *nest* are many. It is not only crazy social media, dictatorship, and disease. It's also our inexperience with a new role in the world and the long lists of emerging disruptions and hardships that might continue to threaten us for some time. The behavior that seems rewarded in the *nest* environment is curiosity, experiment, innovation, and learning. Centuries of economic growth rewarded and honed those skills, too, making another of nature's purposes clear that we will need them more and more as we mature. Looking around, one seems to see an advanced world civilization. The evidence seems to say we are, in fact, now a fully formed but highly undeveloped *new life* on Earth, set to step into a new future, ready or not! Who knows how we will handle it?

# 7 Acknowledgments

---

[19] Henshaw 1990s research archived as "The Physics of Happening" - https://synapse9.com/drwork.htm

# 8   Data Sources

## 8.1   Climate

1.  Atmospheric CO2 PPM 1501-2019                                                    Fig 2
    http://scrippsco2.ucsd.edu/data/atmospheric_co2/icecore_merged_products
    Atmospheric $CO_2$ record from splined ice core data before 1958, and yearly average measurements
    from of Mauna Loa and Antarctica after and including 1958.
    (Keeling & Keeling 2017; Macfarling 2006).

## 8.2   Economic

2.  GDP (PPP) 1971 – 2016*                                                           Fig 2
    Archived IEA PPP data extended with recent World Bank data, see Fig 13 for illustration
    WB: https://data.worldbank.org/indicator/NY.GDP.MKTP.PP.CD?end=2016&start=1990

3.  World economic energy use 1965-2017 –                                            Fig 2
    BP: https://www.bp.com/en/global/corporate/energy-economics/statistical-review-of-world-
    energy/downloads.html

4.  Modern CO2 Emissions – 1971-2016,                                                Fig 2
    Archived IEA CO2 data extended with WRI CO2 emissions: https://www.wri.org/resources/data-
    sets/cait-historical-emissions-data-countries-us-states-unfccc

5.  Historical Co2 Emissions **1751-2013**                                           Fig 2
    US DOE DOE CDIAC data: https://cdiac.ess-dive.lbl.gov/ftp/ndp030/global.1751_2014

6.  World Meat Production – 1961-2016                                                Fig 2
    Rosner - OurWorldInData: https://ourworldindata.org/meat-and-seafood-production-consumption

7.  World Food Production – 1961-2016                                                Fig 2
    FAO: http://www.fao.org/faostat/en/#data/QI